\documentstyle[aps,epsf]{revtex}  

\begin{document}        

\def\inpb{\mbox{$\hbox{pb}^{-1}$}}
\def\Gcs{\hbox{GeV}/\mbox{$c^2$}}
\def\Z{\mbox{$\hbox{Z}$}}
\def\Zstar{\mbox{$\hbox{Z}^*$}}
\def\W{\mbox{$\hbox{W}$}}
\def\WpWm{\mbox{$\W^+\W^-$}}
\def\H{\mbox{$\hbox{H}$}}
\def\epemto{\mbox{$\hbox{e}^+\hbox{e}^- \to$}}
\def\epem{\mbox{$\hbox{e}^+\hbox{e}^-$}}
\def\mpmm{\mbox{$\mu^+\mu^-$}}
\def\tptm{\mbox{$\tau^+\tau^-$}}
\def\lplm{\mbox{$\ell^+\ell^-$}}
\def\nnbar{\mbox{$\nu\bar\nu$}}
\def\qqbar{\mbox{$\hbox{q}\bar{\hbox{q}}$}}
\def\bbbar{\mbox{$\hbox{b}\bar{\hbox{b}}$}}
\def\mH{\mbox{$m_{\hbox{\scriptsize H}}$}}
\def\mZ{\mbox{$m_{\hbox{\scriptsize Z}}$}}
\def\mW{\mbox{$m_{\hbox{\scriptsize W}}$}}

\baselineskip 14pt
\title{Search for the Standard Model Higgs Boson at LEP}
\author{Thomas Greening}
\address{University of Wisconsin -- Madison}
%
\maketitle              

\begin{abstract}        

Preliminary results from the four LEP experiments using data collected at
189\,GeV have shown no evidence for the Standard Model Higgs boson. The
preliminary 95\%
confidence level lower limits on the SM Higgs boson mass from ALEPH, DELPHI,
OPAL, and L3 are 90.4\,\Gcs, 94.1\,\Gcs, 95.5\,\Gcs, and 95.2\,\Gcs,
respectively.
When LEP finishes in the year 2000, each experiment expects to collect
200\,\inpb\ of data at 200\,GeV. These data will allow the discovery of the SM
Higgs boson with a mass lower than about 105\,\Gcs. Assuming that no new
evidence for the SM Higgs boson is found, the mass exclusion limit would be
approximately 110\,\Gcs.

\
\end{abstract}          

\section{Introduction}               

The objectives of my talk are to present current LEP limits on the mass of the
Standard Model (SM) Higgs boson using data from 1998 taken with a
center-of-mass energy of 189\,GeV, and to predict, using all of the
expected data taken by LEP through the year 2000, the discovery potential
and final Higgs boson mass limit, assuming no evidence is found.

Although the Standard Model has had tremendous success in explaining all
known particle physics measurements, the exact mechanism by which the masses
of the vector bosons and fermions are generated is still not understood. The
Glashow-Weinberg-Salam theory describes a mechanism which generates particle
mass through interactions of the particle with a scalar 
field~\cite{greening114ref1}. This Higgs field would manifest itself as a
neutral spin-0 boson called the Higgs boson.

The Glashow-Weinberg-Salam theory predicts all aspects of the Higgs boson,
except for the Higgs boson mass. Electroweak observables do, however,
depend upon the Higgs boson mass through logarythmic corrections. Recent
precision measurements of the top quark mass, $\sin^2\theta_W$, and the
\W\ boson mass, indicate a light Higgs mass near 100\,\Gcs\ and a 95\%
confidence level upper mass limit of 262\,\Gcs~\cite{greening114ref2}.
Although highly uncertain,
this electroweak measurement is exciting since it is at the threshold of the
current mass limits on the Higgs boson.

\section{Higgs Boson Production and Decay}

Since the coupling strength of the Higgs boson to other particles is
proportional to the particle's mass, the Higgs boson is
produced by coupling to heavy particles, of which the heaviest known
particle produced at LEP2 is the \Z\ boson. As a consequence, the main
production mechanism for Higgs bosons at LEP2 is the Higgs-strahlung process
$\epemto\ \Zstar\to\H\Z$ where
the Higgs boson is emitted from the \Z\ boson line~\cite{greening114ref3}. The
other Higgs boson production channels at LEP2 are the \W\W\ and \Z\Z\ fusion
processes which produce a final state with a pair of electron neutrinos or
electrons, respectively. In these fusion processes, the Higgs boson is formed
in the collision of two quasi-real \W\ or \Z\ bosons radiated from the
electron and positron beams.
Interference between the production processes with electron neutrinos
or electrons in the state are taken into account~\cite{greening114ref4}.

\begin{figure}[ht]      
\centerline{\epsfxsize 3.4 truein \epsfbox{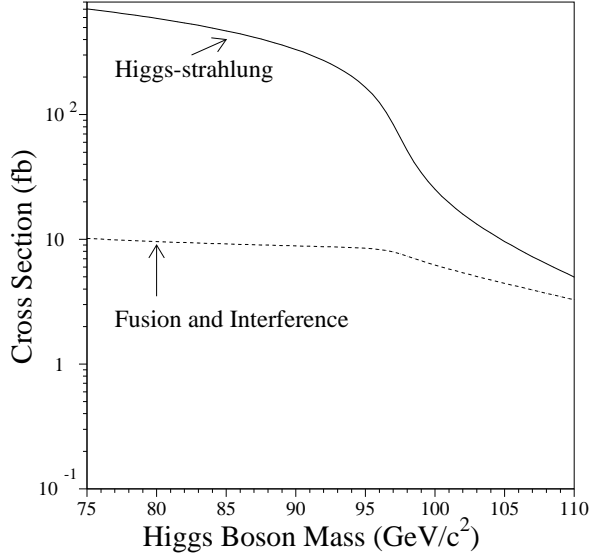}}   
\vskip -0.2 cm
\caption[]{
\label{greening114fig1}
\small
The Higgs boson cross section as a function of Higgs boson mass for the
Higgs-strahlung process, and the sum of the \W\W\ and \Z\Z\ fusion processes
and the interference terms.
}
\end{figure}

The radiatively corrected cross sections for the Higgs-strahlung process and
the sum of the two fusion processes including the interference terms are shown
in Figure~\ref{greening114fig1}
as a function of the Higgs boson mass for a center-of-mass energy of
188.6\,GeV. The rapid fall off in the
cross section for the Higgs-strahlung process at a Higgs boson mass of
95\,\Gcs\ is due to the diminishing phase space available to produce both
the heavy Higgs boson and an on-shell \Z\ boson.

\begin{figure}[htb]      
\centerline{\epsfxsize 3.4 truein \epsfbox{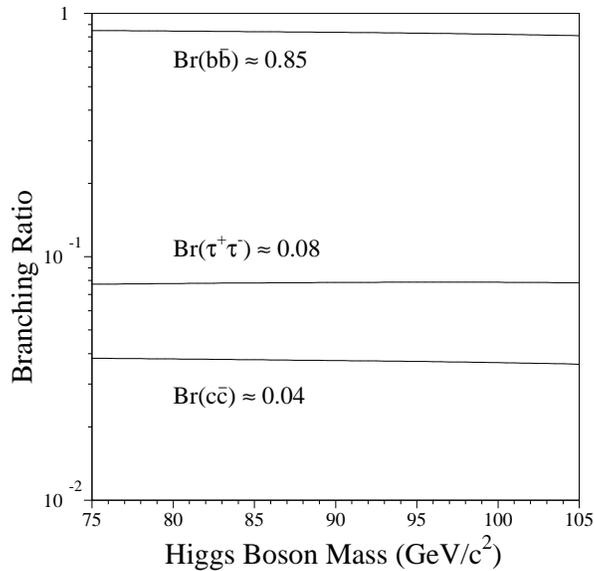}}   
\vskip -0.2 cm
\caption[]{
\label{greening114fig2}
\small 
The branching ratios of the decay of the Higgs boson into b quarks, $\tau$
leptons, and c quarks as a function of the Higgs boson mass.
}
\end{figure}

In the Higgs boson search region of interest from masses of
about 85\,\Gcs\ to 100\,\Gcs,
the decay of the Higgs boson into pairs of top quarks, \Z\ bosons, or
\W\ bosons is not kinematically accessible. Consequently, the Higgs boson,
which couples to mass, will most likely decay into the next heaviest
set of particles which are the b quarks, $\tau$ leptons, and c quarks in
order of decreasing mass. Figure~\ref{greening114fig2} shows the nearly
mass independent branching ratios of the decay of the Higgs boson
as a function of its
mass. The dominant decay to b quarks comprises about 85\% of all Higgs boson
decays, while the decay to $\tau$ leptons provides another 8\% to the total
branching fraction. The searches at LEP2 consider the Higgs boson decaying
to b quark pairs or $\tau$ lepton pairs only, as these two channels comprise
93\% of the total Higgs branching fraction.

\section{Higgs Boson Topologies}

The four LEP collaborations have searched for the Standard Model Higgs boson
in all of the possible decays of the Higgs boson (\bbbar, \tptm) and the
\Z\ boson (\qqbar, \nnbar, \epem, \mpmm, \tptm). Most of the backgrounds with
large cross sections are easily reduced leaving the more difficult
backgrounds which, fortunately, have
cross sections not much larger than the Higgs boson signal. The most
difficult background for all topologies is the \Z\Z\ final state.
This irreducible background occurs when one
\Z\ boson decays to \bbbar\ or \tptm\ and \mH\ is approximately equal to \mZ.
The high purity of the selections and the large amount of
collected luminosity per experiment allows this difficult \Z\Z\ barrier to be
overcome.

\subsection{Four Jet Topology}

The four jet final state, where the Higgs boson decays to b quarks and the
\Z\ boson decays to any quark pair, comprise 64.6\% of the Higgs boson
final states. Difficult backgrounds for this topology include four jet
WW events with four well defined, isolated jets. This background is
significantly reduced by b-tagging the two jets in the event associated
to the Higgs boson. Another difficult background arises from b quark pair
production with a radiated high energy gluon. B-tagging is not effective for
this background, and kinematics must be used to distinguish jets arising
from the \Z\ boson decay in the signal and jets from the high energy gluon.
Typical selection efficiency for this channel is about 40\%. 

\subsection{Missing Energy Topology}

The missing energy topology where the Higgs boson decays to b quarks and
the \Z\ boson decays to neutrinos comprises 20.0\% of the Higgs boson final
states. A difficult background for this topology arises from b quark
pair production with two or more
high energy initial state radiated photons going
undetected down the beam. Cuts on kinematic variables like the acoplanarity
of the b jets and the transverse momentum of the event are used to remove
this difficult background that is both b-tagged and has a large missing
mass. Typical selection
efficiencies for the missing energy channel is about 40\%.

\subsection{Lepton Pair Topology}

The leptonic final state where the \Z\ boson decays to either an electron or
muon pair comprises only 6.7\% of the total Higgs boson final states, but this
final state achieves high purity with the ablility to precisely reconstruct
the \Z\ boson. This  allows the reduction of all backgrounds except the
irreducible \Z\Z\
final state. High purity of the channel also allows sensitivity to the decay
of the Higgs boson to both b quarks and $\tau$ leptons which are typically
reconstructed as the recoil to the lepton pair. Typical selection efficiency
for this channel is typically about 75\%, where the largest losses are due
to the charged particle tracking acceptances of the detectors, and the
insensitivity of the selection to off-shell \Z\ bosons.

\subsection{Tau Pair Topology}

The most difficult channel is the final state containing two jets and a 
pair of $\tau$ leptons. This final state arises from either the decay of
the Higgs boson to b quarks and the \Z\ boson to $\tau$ leptons or the
decay of the Higgs boson to $\tau$ leptons and the \Z\ boson to any
quark pair. The combined branching fraction of these two final states is 8.7\%.
When no b-tagging can be applied, as in the second case were the \Z\ boson
decays into all quark flavors, a difficult background arises from the \WpWm\
process where one \W\ decays to a $\tau$ and a neutrino. Since $\tau$
identification is difficult, another charged particle in the event is often
identified as the decay of the other $\tau$ lepton, and rejection of these
types of events relies upon tight kinematic constraints on the final state
consistent with \H\Z\ production. The typical selection efficiency for this
channel is generally less than 30\%.

\section{Limits on the Higgs Boson Mass}

Both the expected and observed Higgs boson lower mass limits at 95\%
confidence level for each of the LEP experiments are summarized in
Table~\ref{greening114tbl1}. These preliminary results include all
of the data collected at a center-of-mass energy of 189\,GeV and were made
immediately after the end of the physics data taking period of 1998.
Consequently, all of these limits are considered highly preliminary
and likely to change.

\begin{figure}[htb]      
\centerline{\epsfxsize 3.5 truein \epsfbox{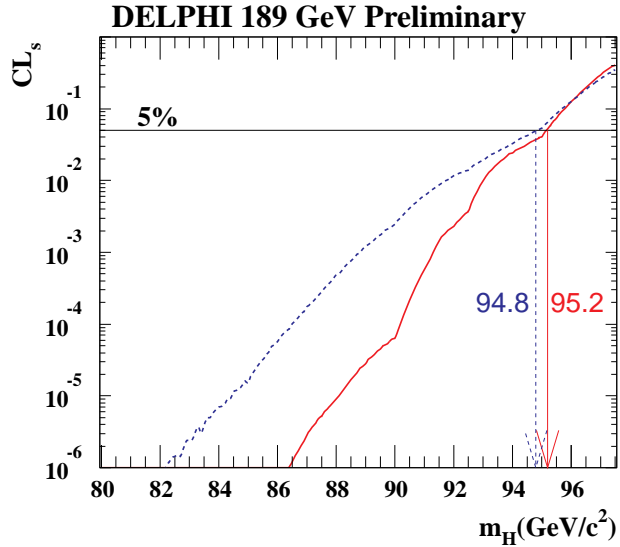}}   
\vskip -0.2 cm
\caption[]{
\label{greening114fig3}
\small The configurations of external legs that are summed over.}
\end{figure}

Figure~\ref{greening114fig3} shows, as an illustration, the preliminary
expected and observed limits from the DELPHI collaboration as a function
of the Higgs boson mass. The intersection of the limits with the 5\% line
define the exclusion region with 95\% confidence. The expected and observed
limits from the four collaborations are in fair agreement indicating the
lack of a signal from the Standard Model Higgs boson.

The similar expected limits for the four collaborations indicate the
similar capabilities of the different detectors. Variations in observed
limits are mostly due to the uncertainties involved in low statistics. The
low ALEPH observed limit could be indication of a Higgs boson signal, but,
considering the limits of the other experiments, the low limit is most likely
due to unluckiness preventing the ALEPH limit from overcoming the \Z\Z\ final
state barrier.

 \begin{table}
 \caption{
\label{greening114tbl1}
Preliminary expected and observed limits on the Standard Model
Higgs boson mass at 189\,GeV.}
 \begin{tabular}{rcc} 
 Experiment&Expected Limit&Observed Limit\\  
 \tableline 
 ALEPH~\cite{greening114ref5}&95.7\,\Gcs&90.2\,\Gcs\\ 
 DELPHI~\cite{greening114ref6}&94.8\,\Gcs&95.2\,\Gcs\\ 
 L3~\cite{greening114ref7}&94.5\,\Gcs&95.5\,\Gcs\\ 
 OPAL~\cite{greening114ref8}&95.2\,\Gcs&94.0\,\Gcs\\ 
 \end{tabular}
 \end{table}

\section{Prospects for the End of LEP}

\begin{figure}[ht]      
\vskip -0.2 cm      
\centerline{\epsfxsize 3.4 truein \epsfbox{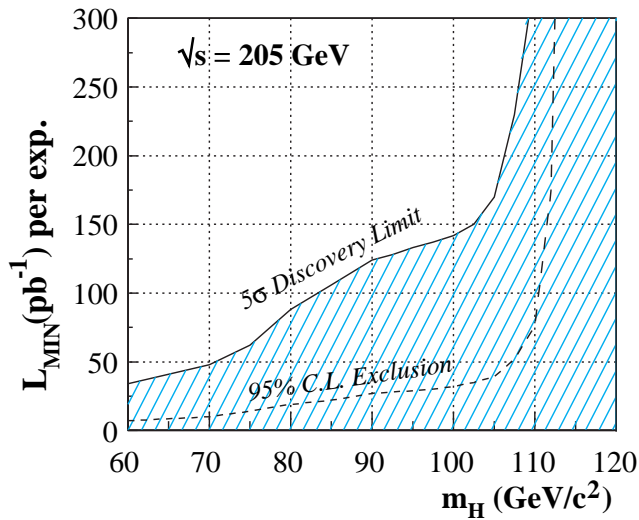}}   
\caption[]{
\label{greening114fig4}
\small Discovery and exclusion potential of the combined LEP experiments
as a function of the collected luminosity per experiment for a center-of-mass
energy of 205\,GeV.}
\end{figure}

By the end of the LEP program in the year 2000, each experiment is expected
to have received about 200\,\inpb\ of data with a center-of-mass energy
of 200\,GeV. Figure~\ref{greening114fig4} shows the discovery potential and
expected limit for the Standard Model Higgs boson as a function of the
luminosity for each experiment~\cite{greening114ref9}.
The figure assumes that the limits from all
four LEP collaborations will be combined.  The figure is an overly
optimistic expectation for LEP performance since it is made assuming
a center-of-mass energy of 205\,GeV which is probably unattainable.  To account
for the higher center-of-mass energy, a few \Gcs\ should be subtracted from
the Higgs boson mass to obtain realistic expectations.  The figure indicates
that with 200\,\inpb\ of data per experiment, LEP2 should be able to discover
a Standard Model Higgs boson with a mass less than about 105\,\Gcs. Assuming
no new evidence for the Higgs boson, exclusion limits at 95\% confidence level
will set a minimum mass on the Standard Model Higgs boson at about 110\,\Gcs.

\end{document}